\documentclass[aps,prd,preprintnumbers,
floatfix,nofootinbib, twocolumn]{revtex4}

\usepackage{graphicx,amssymb,amstext,mathrsfs,dsfont,amsfonts}
\usepackage{wrapfig,boxedminipage,setspace,subfigure,epsfig}
\usepackage{amsxtra,amsmath}
\usepackage{color,latexsym,url}
\usepackage[colorlinks
,hyperindex]{hyperref}
    \hypersetup
    {
        colorlinks,%
        citecolor=red,%
        linkcolor=blue,%
        urlcolor=black,%
    }

\def\a  {\alpha}       \def\b  {\beta}         
       \def\d  {\delta}        
\def\e  {\epsilon}        
\def\l  {\lambda}             \def\m  {\mu}
\def\n  {\nu}                  
                 
   \def\w  {\omega}

 \newcommand{\call}{\mbox{${\cal L}$}}

\def\IR{{\hbox{{\rm I}\kern-.2em\hbox{\rm R}}}}
\def\IB{{\hbox{{\rm I}\kern-.2em\hbox{\rm B}}}}
\def\IN{{\hbox{{\rm I}\kern-.2em\hbox{\rm N}}}}
\def\IC{\,\,{\hbox{{\rm I}\kern-.59em\hbox{\bf C}}}}
\def\IZ{{\hbox{{\rm Z}\kern-.4em\hbox{\rm Z}}}}
\def\IP{{\hbox{{\rm I}\kern-.2em\hbox{\rm P}}}}
\def\IH{{\hbox{{\rm I}\kern-.4em\hbox{\rm H}}}}
\def\ID{{\hbox{{\rm I}\kern-.2em\hbox{\rm D}}}}

\def\be{\begin{equation}}
\def\ee{\end{equation}}
\def\ba{\begin{eqnarray}}
\def\ea{\end{eqnarray}}

\def\half{\frac{1}{2}}

\def\ra{\rightarrow}

\def\del{\partial}

\def\det{{\rm det}}

\def\nn{\nonumber}
\def\ea{{\it et al}. }

\newcommand{\wt}{\widetilde}

\newcommand{\beq}{\begin{equation}}
\newcommand{\eeq}{\end{equation}}
\newcommand{\bea}{\begin{eqnarray}}
\newcommand{\eea}{\end{eqnarray}}

\newcommand{\guv}{{g_{\mathrm{UV}}}}
\newcommand{\guvt}{{g^2_{\mathrm{UV}}}}

\begin{document}

\newcommand\sect[1]{\emph{#1}---}

\preprint{
\begin{minipage}[t]{3in}
\begin{flushright} SHEP-12-21
\\[30pt]
\hphantom{.}
\end{flushright}
\end{minipage}
}

\title{Inflation from Strongly Coupled Gauge Dynamics}

\author{Nick Evans}
\email{evans@soton.ac.uk}
\affiliation{ School of Physics and Astronomy, University of
Southampton, Southampton, SO17 1BJ, UK  \vspace{-0.2cm}
}
\author{James French}
\email{james.french@worc.oxon.org}
\affiliation{ School of Physics and Astronomy, University of
Southampton, Southampton, SO17 1BJ, UK   \vspace{-0.2cm}
}
\author{Keun-Young Kim}
\email{K.Y.Kim@uva.nl}
\affiliation{ Institute for Theoretical Physics, University of Amsterdam, Science Park 904, \\ Postbus 94485, 1090 GL Amsterdam, The Netherlands
\vspace{0.2cm}}


\begin{abstract}
We argue, using a phenomenological holographic approach, that walking, strongly coupled gauge theories
generate a suitable potential for a small field inflation model. We show that the effective description is 
a model of a single inflaton. We determine the tunings necessary in the gauge sector to generate inflation and
overcome the $\eta$ problem.

\end{abstract}

\maketitle


Inflation (see for example \cite{Lyth:1998xn}) is a key ingredient in cosmology. The usual description has
a scalar field slow rolling in some suitably designed potential. Scalar fields are though
technically unnatural (in the absence of supersymmetry, and putting aside the possible recent discovery of a fundamental higgs field!) but it is possible that they
are an effective description of some strongly coupled symmetry breaking sector. Historically it
has not been possible to directly compute in such a strongly coupled theory but the AdS/CFT Correspondence 
\cite{Malda,Witten:1998qj,Witten:1998b,Gubser:1998bc} does now allow such study. 

In this paper we wish to return to the study of a model \cite{Evans:2010tf} in which the inflaton is a quark anti-quark ($\bar{q}q$)
bound state in a strongly coupled system\footnote{Some other recent related discussions can be found in \cite{Chen:2010pd,Channuie:2011rq,Bezrukov:2011mv}}. The field theory is in the spirit of QCD, breaking the quark's
chiral symmetry by the dynamical generation of a $\langle \bar{q} q \rangle$ vev, although we imagine this
dynamics occurs at a much higher scale than in QCD.  We consider some asymptotically free gauge theory that runs
from a perturbative UV to a strongly coupled IR regime. 
An effective description of the strongly
coupled IR will be provided by a holographic construction in the spirit of AdS/QCD \cite{Erlich:2005qh,Da Rold:2005zs}
based on the D3/D7 system \cite{Polchinski,Bertolini:2001qa,Karch,Mateos,Erdmenger:2007cm}. 
The asymptotically free UV is replaced in holographic descriptions by a conformal but still strongly
coupled regime.
Crucially the model  includes the dynamics of chiral symmetry breaking \cite{Babington,Ghoroku:2004sp,Kutasov:2012uq,Alvares:2009hv}. We parameterize the running
of the gauge coupling in the theory to allow us to study theories that range from near to far from
conformal. For some choices of parameters we can engineer a very flat effective potential for the quark condensate. 
Inflation can then be obtained by placing the condensate near zero and allowing it to slow roll to the true vacuum. 
In the holographic dual the D7 brane rolls from a flat to a curved configuration in an AdS geometry with a particular
choice of dilaton.

Naively the motion of the D7 brane, which is extended in the holographic radial direction of AdS, appears to 
correspond to the dynamics of a large number of coupled inflaton fields. In our previous analysis \cite{Evans:2010tf} of the model we therefore resorted
to a numerical computation within the DBI action for the D7 brane of the time dependent dynamics of the roll. We then
numerically searched for configurations that led to longer roll times. In this paper we can present a much more clear cut understanding of the model. 
In particular we understand that the tuning of parameters that we used was tuning us
onto the chiral restoration transition analagous to that that occurs at the edge of the conformal window in QCD 
\cite{Appelquist:1996dq,Appelquist:1998rb,Dietrich:2006cm} leading
to walking gauge dynamics \cite{Holdom:1981rm}. This type of transition has been explored in detail using holography in 
\cite{Jarvinen:2009fe,Antipin:2009dz,Jarvinen:2011qe,Kutasov:2012uq,Alvares:2012kr}. 
Further here we will show that in fact
just a single eigenmode of the D7 brane, ie a single mesonic state, is playing the role of the inflaton. The model is    
formally equivalent to a single inflaton model. The cosmological phenomenology is then
already well known \cite{Chen:2010xka}. We will also address the $\eta$ problem \cite{Copeland:1994vg} in the model.

It is important to stress that we will not solve any of the usual fine tunings needed in traditional inflaton models.
Our aim here is simply to recast the problem for strongly coupled gauge theories and understand what those tunings imply
for the running of such theories.

\section{The holographic description}

\begin{figure}[]
\centering
   {\includegraphics[width=6cm]{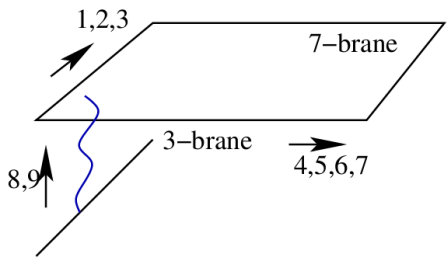}}
\caption{A sketch of the D3/D7 construction that our model is based on.}\label{D3D7}
\end{figure}

The holographic description is based on the simplest brane
construction of a 3+1d gauge theory with quarks which is
the D3/D7 system ($N_f$ number of probe D7 branes in the background of $N_c$ number of D3 branes)\cite{Polchinski,Bertolini:2001qa,Karch,Mateos,Erdmenger:2007cm}. See Figure \ref{D3D7} for the branes set-up. 
The basic gauge theory is large $N_c$, ${\cal N} = 4$ super Yang-Mills with $N_f$ quark fields.
We will work in the quenched approximation ($N_c \gg N_f$) which in the gravity dual corresponds to the probe approximation
\cite{Karch}. The theory has a $U(1)$ R-symmetry under which a
fermionic quark anti-quark condensate has charge 2 and
plays the role of $U(1)$ axial \cite{Babington}.

Top down models of this type exist with chiral
symmetry breaking\footnote{See \cite{Evans:2012cx,Evans:2011eu,Evans:2011tk} for related studies of the
temperature-chemical potential chiral (axial) phase structure of some of the models we study.}. Supergravity solutions are known that
correspond to the AdS space being deformed by a running coupling 
introduced by a non-trivial dilaton \cite{Babington,Ghoroku:2004sp}. In cases where the coupling grows in the
IR chiral
symmetry breaking is induced. These models have very
specific forms for the running coupling and are typically
singular somewhere in the interior. 
A simpler and completely computable case with
chiral symmetry breaking is provided by introducing a
background magnetic field associated with $U(1)$ baryon
number \cite{Magnetic,Evans:2010iy}. At the level of the DBI action the magnetic field term can be considered as a 
non-back reacted dilaton profile.

Here we will work more phenomenologically introducing a dilaton
profile to induce symmetry breaking but without back reacting it
on the AdS geometry \cite{Alvares:2009hv}. The idea is to work in the spirit of AdS/QCD \cite{Erlich:2005qh,Da Rold:2005zs}
and use different dilaton profiles to study gauge theories with
different IR running.  i.e. we will allow a dilaton 
to represent the running of the gauge theory coupling
\beq
e^{\Phi} = g^2_{\mathrm{YM}}(r) = \guvt ~\beta(r) \ , \label{beta}
\eeq
where the function $\beta(r)$ will be specified in the next section with $\beta(r) \rightarrow 1$ as
$r \rightarrow \infty$ ($r$ is defined in \eqref{ametric}).

We will take a background holographic geometry AdS$_5$$\times S_5$
\begin{eqnarray}
  && G = {1 \over \guv} \left[{r^2
\over R^2} ( g_{tt}dt^2 + g_{ij} dx^idx^j) \right. \nn \\
 && \quad \left. + {R^2 \over r^2} \left( d\rho^2 + \rho^2 d\Omega_3^2 + dw_5^2 +
dw_6^2 \right)\right] \ , \label{ametric}
\end{eqnarray}
where $R^4 = 4 \pi \guvt N_c \alpha'^{2}$ and the second line parameterizes $\mathbb{R}^6$, where $\{\rho, \Omega_3\}$ correspond to $x^4$-$x^7$ and $w_{5(6)}$ corresponds to $x^{8(9)}$ in Figure \ref{D3D7}.  The metric $g_{tt} = -1$ and $g_{ij} = a(t)^2 \d_{ij}$, where $a(t)$ is the expansion factor reflecting 
inflation.  Formally we should ask how the
AdS aspects of the geometry respond to the presence of the inflation but we will
leave discussion of such back reaction until the final section. As usual the radial 
direction $r= \sqrt{\rho^2 + w_5^2 + w_6^2}$ is the energy scale of the gauge theory.

\begin{figure}[]
\centering
   {\includegraphics[width=7cm]{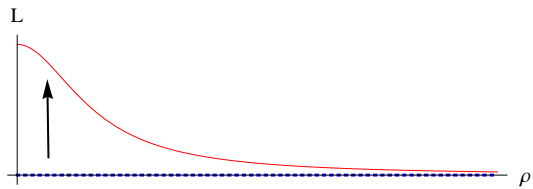}}
\caption{  Evolution of the D7 brane embedding from the false vacuum ($L=0$, dotted blue line) to true vacuum ($L \ne 0$, solid red curve).    }\label{embedding}
\end{figure}

Quarks are introduced through a probe D7 brane (Fig \ref{D3D7}). 
The DBI action of the D7 brane in Einstein frame is given by
\beq \label{DBI}
 S_{DBI} = -T_7 \int d^8\xi e^\Phi  \sqrt{- \det P[G]_{ab}}
\eeq
where $T_7 = 1/(2 \pi)^7\alpha'^{4}$ and $P[G]_{ab}$ is the pull back of the background metric \eqref{ametric} onto the D7. By assuming $L = \w_5(t, \vec{x},\rho)$ with $\w_6 = 0$, we have 
\beq \label{DBI}
 S_{DBI} = \int d^4x \sqrt{-g} \, \call \,,
\eeq
where
\begin{eqnarray}
&&\frac{\mathcal{L}}{\wt{T}_7 R^4 } \equiv - \int d \rho \, \b \,
\frac{\rho^3}{\rho^2 + L^2} \times \nn \\
&&\qquad \sqrt{ {g^{\m\n} \del_\mu{L}\del_\nu{L} + (\rho^2 + L^2)^2(1+L'^2) }} \,. \label{Lag4D}
\end{eqnarray}
$\wt{T}_7 = 2 \pi^2 T_7/ \guvt$ and $L' = \del_\rho L$.
All variables ($\rho, L$, and below $\l $) in the integral (\ref{Lag4D}) are rescaled by $R$ and
dimensionless.
For the moment we leave the dilaton as an unspecified function that asymptotes to unity at large $r=\sqrt{\rho^2 + L^2}$.
We will consider several cases below.

Typical $t,x$ independent, vacuum configurations are easily found numerically by
solving the appropriate Euler-Lagrange equation for the D7 profile $L(\rho)$.  Solutions corresponding to massless quarks fall off asymptotically as 
\begin{equation} \label{asymp}
  L \sim \frac{c}{\rho^2} \,,
\end{equation}
with $c$ a measure of the quark condensate $\langle \bar{q} q \rangle$ and satisfy $L'(0)=0$ \cite{Babington}. There is always the symmetric $L=0$ solution  but
usually symmetry breaking configurations exist that curve off the $L=0$ axis as well.  See Fig \ref{embedding}. By computing the on shell actions for both cases, we may identify which solution corresponds to the true vacuum. In this paper, we only consider the case the embedding of $L \ne 0$ corresponds to the true vacuum. 

Therefore, as shown by the arrow in Figure \ref{embedding}, any initial embedding has to ``roll down'' towards the red solid curve as time goes on. This embedding dynamics is the holographic image of our field theory composite inflation dynamics\footnote{This time-dependent dynamics also has been studied in \cite{Evans:2010xs} in the context of cooling quark-gluon plasma in RHIC.}. The time dependent composite inflaton may be 
identified with the quark condensate $c(t) = \langle \bar{q} q \rangle (t)$ in \eqref{asymp}.

\section{A Flat Inflaton Potential}

Our first task is to use the holographic model to find dilaton/running coupling profiles that generate a 
flat potential that can be used for a small field inflaton model.

In \cite{Evans:2010tf} we used the step-function form for the dilaton
\begin{eqnarray}
 \beta =  A +1 - A \tanh\left[ (r - \lambda) \right] \ , \label{beta}
\end{eqnarray}
Here $\lambda$ is the intrinsic scale at
which the conformal symmetry is broken and corresponds loosely to $\Lambda_{QCD}$ in QCD. $A$ controls
the step height. 

Previously \cite{Alvares:2012kr} we analyzed the phase structure of such theories as a function of A. We studied the linearized action
about the flat chiral symmetry preserving configuration $L=0$, $L'=0$. If we consider $x,t$ independent
configurations in (\ref{Lag4D}) then we can make the change of variables 
\begin{equation}
\label{rt} \tilde{\rho} = \sqrt{{1 \over 2} { 1 \over \int_\rho^\infty {d\rho \over \beta \rho^3}} }  \,,
\end{equation}
If we write $L = \tilde{\rho} \phi$ then the action for the scalar $\phi$ is that of a canonical scalar in AdS with mass given by 
\begin{equation} \label{deltam}
m^2 = -3  - \d m^2 \,, \qquad  \d m^2 \equiv - \beta  {\rho^5 \over \tilde{\rho}^4  } {d \beta  \over d \rho} \,.
\end{equation}
Clearly our step function form for $\beta$ gives a negative contribution to the mass squared of the scalar during the 
range of the step. If the mass falls below the Breitenlohner-Freedman (BF) bound \cite{Breitenlohner:1982jf}
in AdS$_5$ of $m^2=-4$ then an instability exists. Clearly
there is a critical $A$ needed to violate the BF bound. When,
as is the case for this ansatz for $\beta$, the instability exists at an intermediate range of $\rho$ 
the transition occurs when the BF bound is violated over a sufficiently large range in $\rho$ and the transition
is second order in nature. 

Indeed for this ansatz if one performs the full numerical analysis one finds a second order transition from
the flat embedding to a curved chiral symmetry breaking embedding at $A=0.8$. Of course as one approaches this transition
the difference in energy between the flat and curved embeddings, $\Delta V$, falls to zero. Similarly the quark condensate,
$\langle \bar{q} q \rangle$, also falls towards zero at the transition point. The numerical analysis of \cite{Evans:2010tf} though 
found that as one approached the transition, rescaling $\lambda$ to keep $\Delta V$ constant, inflationary
behaviour resulted. Note that keeping $\Delta V$ fixed is the appropriate way to compare inflating theories - we want
the vacuum energy of the theories, which controls the expansion rate, to be the same in each case. 
Now we understand we are tuning to the phase transition we can make this more concrete. 
To achieve a very flat potential we need that the dimensionless quantity $\langle \bar{q} q \rangle / \Delta V^{3/4}$
grow as we tune to the transition point. If this is the case then the fixed drop in the potential $\Delta V$ occurs over
a much larger range in the condensate and the potential becomes very flat for condensate values below the minimum.

\begin{figure}[]
\centering
\includegraphics[width=6.5cm]{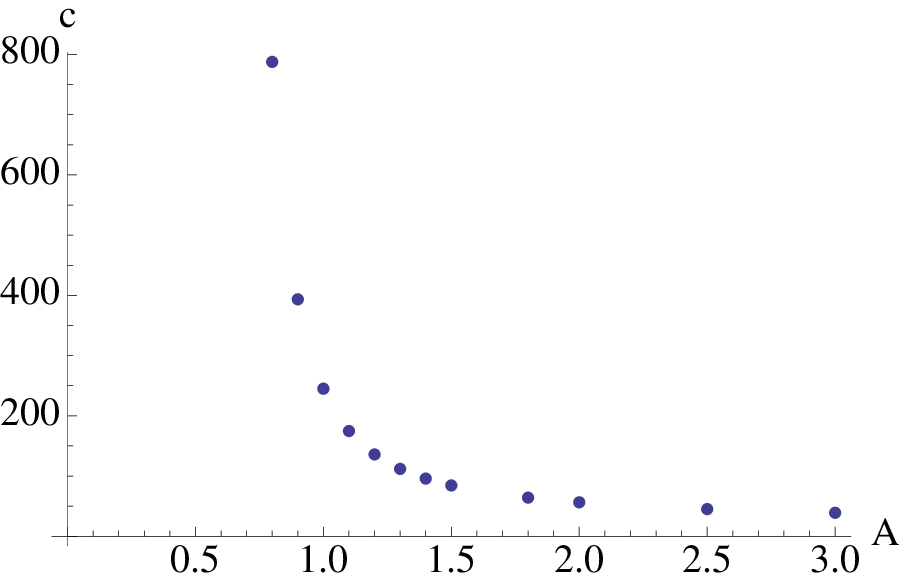}
\includegraphics[width=6.5cm]{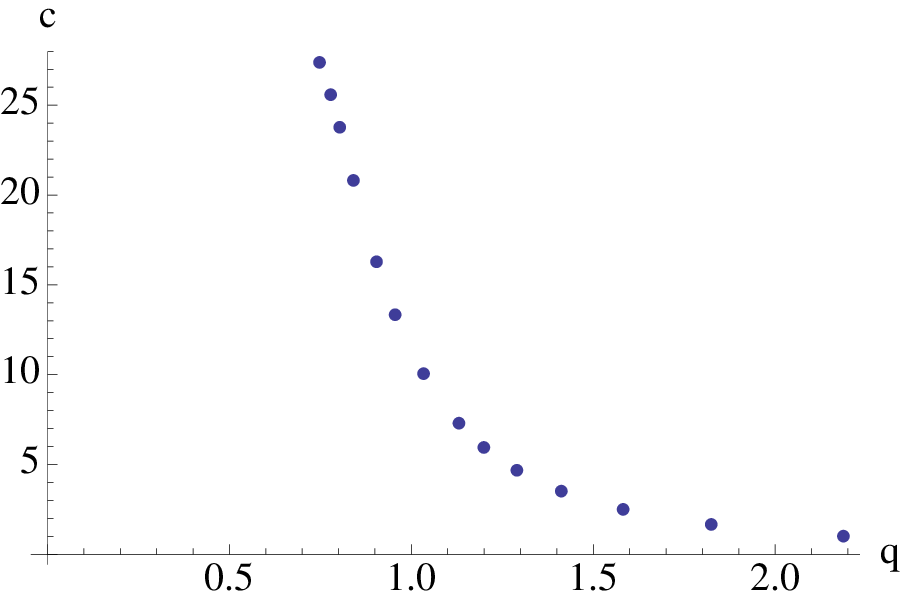}
\caption{The quark condensate $\langle \bar{q} q \rangle$ against $A$ in the model of (\ref{beta}) and $q$ in (\ref{beta2}) in each case for fixed vacuum energy $\Delta V$.}
\label{potshape}
\end{figure}

In Fig \ref{potshape}a we plot  $\langle \bar{q} q \rangle$ against $A$, after tuning $\lambda$ so that $\Delta V$ is the 
same in each case, showing that it indeed diverges as we approach
the critical value of $A$. This explains why the model makes a good inflaton model.

In \cite{Evans:2010tf} we argued that since minimizing the step size $A$ led to inflation this suggested walking dynamics in a gauge theory
might also. The holographic understanding of walking has greatly improved recently  - see \cite{Jarvinen:2009fe,Antipin:2009dz,Jarvinen:2011qe,Kutasov:2012uq,Alvares:2012kr}. The usual picture for a walking gauge theory is a theory lying close to the edge of the ``conformal window'' 
\cite{Appelquist:1996dq,Appelquist:1998rb,Dietrich:2006cm}.
The conformal window in QCD is a range in the number of quark flavours over which the theory is expected to be asymptotically 
free but approach an IR conformal fixed point.  At the lower edge in the number of flavours it is 
presumed that the coupling value at the fixed point is sufficiently large to trigger chiral symmetry breaking.  
In perturbative QCD the anomalous dimension of the quark condensate is proportional to the coupling value. Schwinger Dyson analysis suggests that chiral symmetry breaking sets in when the dimension of the condensate falls to about 2 \cite{Appelquist:1988yc,Cohen:1988sq,Miransky:1988gk}. 
In the holographic analysis the scalar $\phi$ we introduced below  (\ref{rt}) is dual to the quark condensate. The naive AdS dictionary relates its mass to the dimension of the quark condensate, $\Delta$, according to $m^2 = \Delta (\Delta-4)$.
The BF bound for instability therefore again corresponds to the condensate dimension falling to 2. 

The model above therefore confirms some of the the normal intuition for the onset of the chiral transition. However, the ansatz for
$\beta$ in (\ref{beta}) does not lead to an IR fixed point for the scalar mass squared or equivalently the condensate's anomalous dimension. Instead the extra contribution to the scalar mass (\ref{deltam}) is only present across the step in $\beta$ where the derivative is non-zero. In \cite{Alvares:2012kr} we presented a simple fix for this aspect of the quark physics which hopefully makes the dynamics closer to that of walking theories. If we set $\beta \sim 1/ \rho^q$ in the IR then we find
\begin{equation}
\delta m^2 = {4 q \over (2 - q)^2} \,.
\end{equation}
This form for $\beta$ ensures a fixed IR shift in the mass squared. Varying $q$ so that $m^2$ moves through -4 leads to a chiral phase transition 
which shows Miransky scaling \cite{Miransky:1996pd} or BKT-like behaviour \cite{Kaplan:2009kr,bkt,Evans:2010hi} (the condensate grows exponentially from
the transition point). This behaviour is that expected at the walking edge of the conformal window.

To make a consistent model we must find a form for $\beta$ that has the required IR form but asymptotes to unity at large AdS radius. The simplest example is
\begin{equation} \label{beta2}
\beta = 1+{B \over r^q}  = 1+ { B \over (\rho^2 + L^2)^{q/2} } \,.
\end{equation} 
It is important to stress that the chiral transition is triggered by the far IR behaviour of the ansatz and the UV extension is largely unimportant. 

The choice of $\beta$ in (\ref{beta2}) therefore provides a decent stab at describing the physics of the BKT transition at the edge of the conformal window. Does our previous supposition that these theories will generate a very flat potential suitable for inflation still hold? It is straight forward to check - we can solve the full embedding equation numerically 
and plot $\langle \bar{q} q \rangle$ against $q$, after tuning $B$ so that $\Delta V$ is the same in each case. We show 
this plot in Fig \ref{potshape}. We indeed again see that the condensate diverges and hence the potential becomes 
flat as we tune to the critical $q=0.536$. Walking dynamics and inflation again seemed to be linked in this framework.

\section{The Roll Dynamics}

In \cite{Evans:2010tf}  we studied the time dependent roll of the symmetric flat D7 embedding to the symmetry breaking solution
in the background of the dilaton profile in (\ref{beta}).
We showed numerically that as the parameters approached the critical coupling described above, at fixed $\Delta V$, the roll time grew, supporting
the idea that such theories will generate inflation. To set the roll moving we used an initial
velocity distribution
\begin{equation} \dot{L}(\rho, t = 0) = ve^{-\rho^2} \,. \end{equation}
We found the time to reach the vacuum state was largely independent of this form
although of course it does depend on the total energy injected.

Naively this model looked to us to be a multi-inflaton problem in 3+1d because of the $\rho$ dependence
in the D7 profile $L(\rho,t)$ in the holographic description. Here though we will show that the model is equivalent to a single
inflaton model.

\subsection{Linearized Regime Analysis}

At early times the brane lies close to
$L(\rho)=0$ and has small $L'(\rho)$ so it is natural to consider a linearized regime.
The quadratic Lagrangian $\call_{(0)}$ reads:
\begin{eqnarray}
  && \call_{(0)} =  \int d\rho\ a(t)^3 \left(
  \frac{\b_0}{2\rho} \dot{X}^2 - \frac{\b_0 \rho^3}{2} X'^2 - \frac{\b_0' \rho^2}{2} X^2 \right) \,, \nn \\ \label{Lag0}
\end{eqnarray}
where
\begin{eqnarray}
  &&\b_0 := A+1-A \tanh\left( \rho-\l \right) \,, \nn \\
  &&\b_0' = -A\, \mathrm{sech}^2  \left(\rho-\l \right) \,.
\end{eqnarray}
$X = \delta L$ denotes the linear fluctuation around $L=0$.
This expansion is not completely legitimate (and this is why we had not pursued  it in the previous
paper).
The expansion may break down as $\rho \ra 0$ in Eq.(\ref{Lag0})
since $X$ may not be smaller than $\rho$.
Thus we are assuming either 1) $X(0) \ra 0$ faster than $\rho \ra 0$ at least for
the slow rolling regime or 2) there is a
IR cut-off $\rho_{c}$ so that $\rho_{IR} \gg X(\rho_{c})$.
We will adopt the second and introduce an IR cut-off, $\rho_{c} \equiv \e$ - the justification for
this will be retrospective based on its success as we discuss in detail below.

The linearized equation of motion is
\begin{eqnarray}
  -\frac{\b_0}{\rho} \frac{\del_t(a^3 \dot{X})}{a^3} =  -(\b_0 \rho^3 X')' + \rho^2\b_0'X  \,. \label{EOM1}
\end{eqnarray}
We can separate variables with the ansatz  $X(t,\rho) \equiv T(t)R(\rho)$
and
the equation of motion reads
\begin{eqnarray}
  && \ddot{T} + 3 H \dot{T} + m^2 T = 0 \ , \label{EOMT}  \\
  && R'' + \left( \frac{3}{\rho} + \frac{\b_0'}{\b_0}\right) R'
  + \left(- \frac{\b_0'}{\rho \b_0} + \frac{m^2}{\rho^4} \right) R = 0 \,, \label{EOMR}
\end{eqnarray}
where $m^2$ is an arbitrary separation constant for now, but in principle will be
determined by the embedding dynamics along the $\rho$-direction in the background of $\b$.
So $m^2$ is implicitly a function of $A, \l$ and can be written as
\begin{eqnarray}
  m^2 &=& \frac{\rho}{\b_0} \left(-\frac{(\b_0 \rho^3 R')'}{R}  + \rho^2 \b_0'  \right) \label{m1} \\
      &=& \frac{\int d \rho \left(\b_0 \rho^3 R'^2 + \b_0' \rho^2 R^2\right)}{\int d \rho \frac{\b_0}{\rho} R^2} \,, \label{m2}
\end{eqnarray}
where the first condition comes from separation of variables and the second
is derived from the first and also can be understood as a completeness relations for the self-adjoint operator.
Indeed the Sturm-Liouville theorem applies to (\ref{EOMR}).
\begin{eqnarray}
  \hat{O} R(\rho) =   m^2 w(\rho) R(\rho) \,,
\end{eqnarray}
where
\begin{eqnarray}
  &&\hat{O}R = (-\rho^3 \b_0 R')' + \rho^2\b_0' R, \hspace{0.5cm} w(\rho) = \frac{\b_0}{\rho} \,.
\end{eqnarray}
The natural way to proceed to the solutions is as follows: firstly one could solve (\ref{EOMR})
with $R \rightarrow 0$ at large $\rho$ and subject to $R'(0)=0$. One would expect a tower
of solutions corresponding to discrete eigenvalues $m^2$. One could then
solve for $T(t)$ using (\ref{EOMT}). The solution of that equation
is simply
\begin{equation} \label{Tsol}
T(t) = e^{\alpha t}, \hspace{1cm} \alpha^2 + 3 H \alpha + m^2 = 0 \,.
\end{equation}
$\alpha$ is positive only if $m^2 < 0$. Thus after some evolution only the unstable
mode (or potentially modes) would remain in the solution.

We can not complete the program above in the linearized regime because
the linearized approximation is not valid at
small $\rho$. To circumvent this one has to return
to the full solutions to show that the linearized approximation correctly captures
the main physics of the D7's roll and to compute $m^2$.

\begin{figure}[]
\centering
\includegraphics[width=6.5cm]{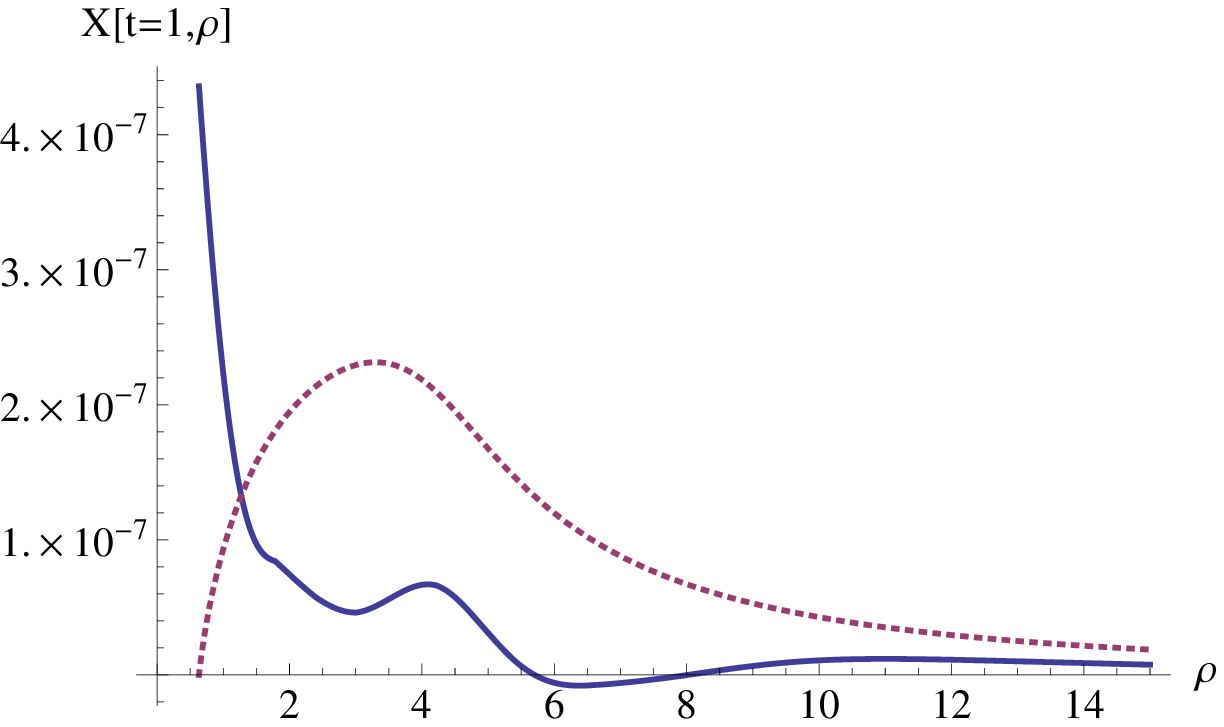}
\includegraphics[width=6.5cm]{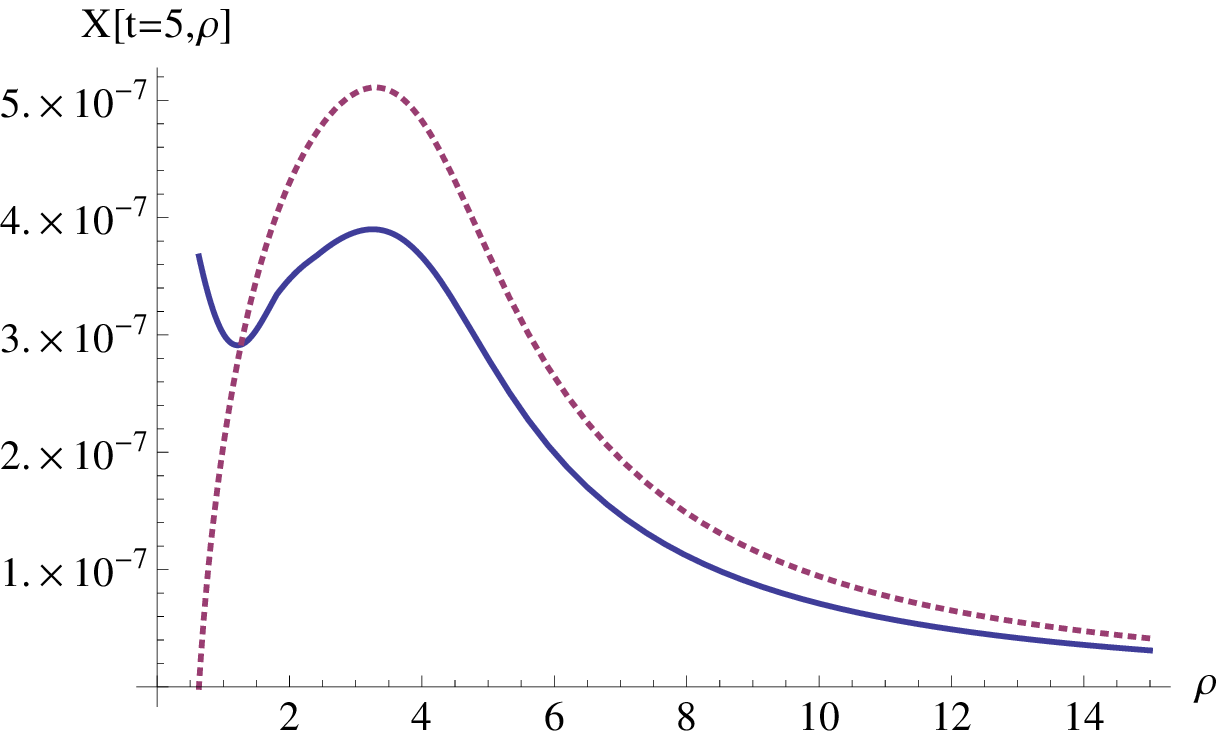}
\includegraphics[width=6.5cm]{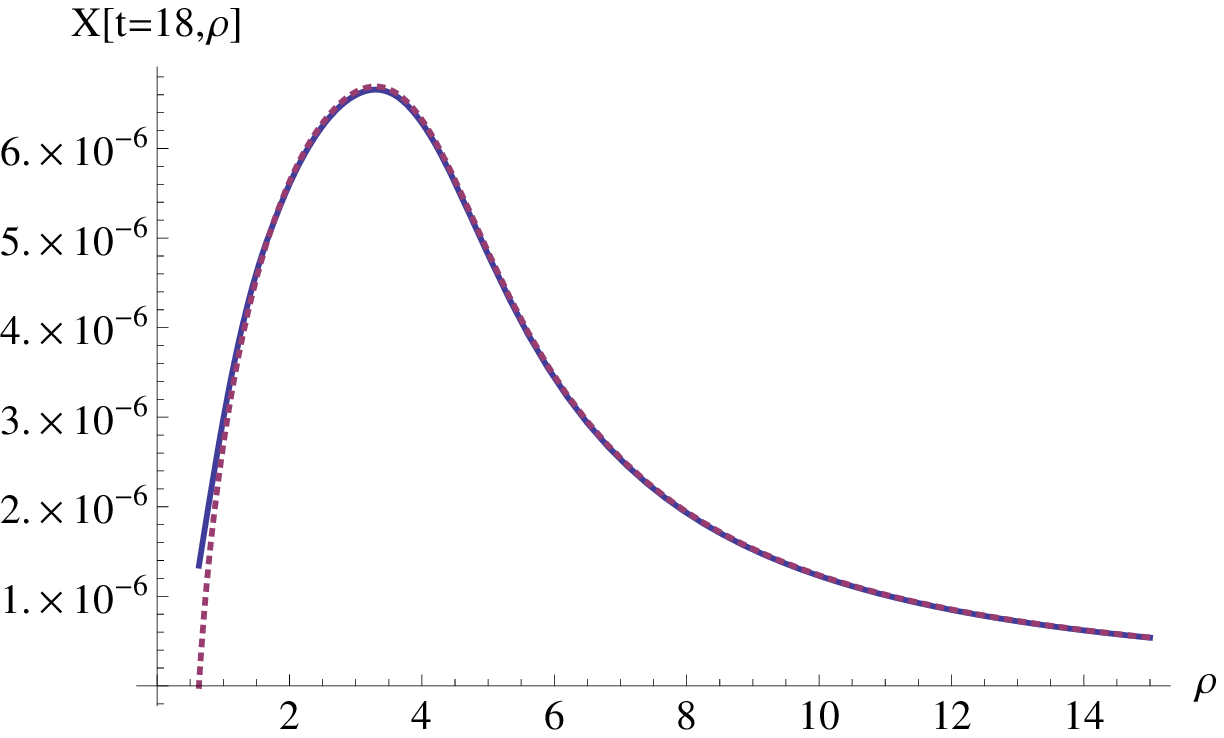}
\includegraphics[width=6cm]{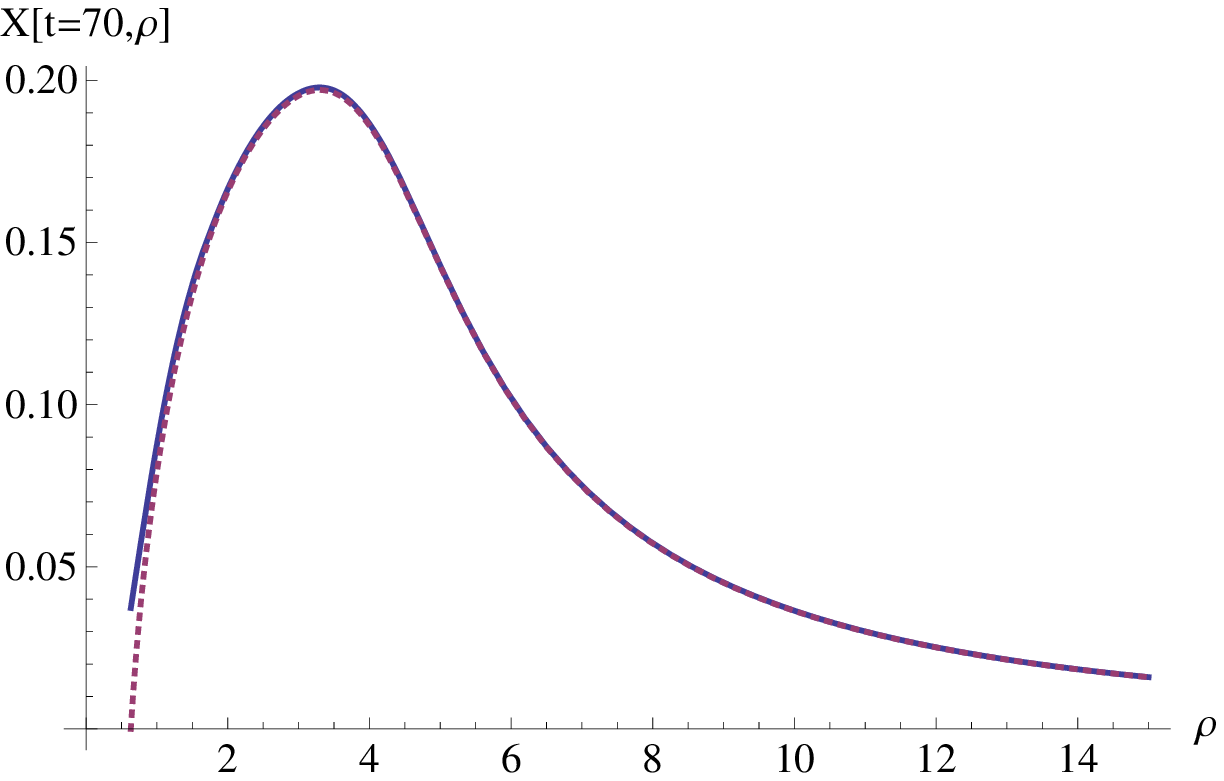}
\caption{Evolution of full numerical embedding at different times (blue), compared with linearised solution (purple).}
\label{profiles}
\end{figure}

Let us show this procedure in a specific example.
Consider the parameter set used in \cite{Evans:2010tf} $A=3,  \l=3.24$.
The evolution of the embedding at times
$t=1$, 5, 18 and 70 is shown in Fig. \ref{profiles}.
From $t=5$ onwards the $\rho$ profile of the solution stabilizes (initially at $\rho>1$) which fits with a separation
of variables in the solution.

Assuming that separation of variables,
the function $T(t)$ may be obtained from examining the time dependence of any one point on the embedding (eg $X(t,\lambda)$). The solution to (\ref{EOMT}) is $T(t)\sim \exp(\alpha t)$. If the linearised approximation is valid, the time-derivative of the lograrithm of $T(t)$ should be a constant, $\alpha$, and that is indeed what we find.
We can now determine $m^2$ for the most unstable mode by simply reading off the value of
$\alpha$ from  the exponential decay for this point.
In this specific case, $\a = 0.198$ and $m^2 = -2.907$.

We can now perform a further consistency check. We can take this value of $m^2$ and
plug it into (\ref{EOMR}). We then solve (\ref{EOMR}) for $R(\rho)$ numerically by shooting in from
large $\rho$. By hand we cut off the solution in the IR at the point where $R(\rho)=0$ and assume that below
this $\rho_{c}$ the linearized solution is not valid but also that the physics below this small $\rho$ value
is not crucial to the dynamics. For example in our specific case we find
\begin{eqnarray}
  X = 4.47\times10^{-7} e^{0.198 t} R(\rho) \ ,   \label{X1}
\end{eqnarray}
where $R(\rho)$ is normalized as $ \int^\infty_{\rho_c} w R^2 = 1$. 
The pre-factor is fixed by matching to the full numerics at $t= 18$.
We now plot these solutions (\ref{X1}) as the dotted curves in Figure \ref{profiles}.
Beyond $t=17$ we have captured the full evolution very well within the linearized approximation.
In particular this confirms that there is a single unstable mode that is dominating the
evolution.

The embedding matches the linearized evolution until around $t=80$ which is essentially
the full roll period.  We also note that the peak in $R(\rho)$
is centred at $\rho=3.24$ which, as expected,
is near the point $\rho=\lambda=3$  where the step change in the dilaton is found.

As a final numerical consistency check, we can evaluate $m^2$ from (\ref{m1}) and (\ref{m2}) numerically
using our $R(\rho)$ solution. For example
\begin{eqnarray}
  m^2 = \frac{\int_{\rho_c}^\infty d \rho \left(\b_0 \rho^3 R'^2 + \b_0' \rho^2 R^2\right)}{\int_{\rho_c}^\infty d \rho \frac{\b_0}{\rho} R^2} =-2.907
\end{eqnarray}
from (\ref{m2}).
The consistency is very satisfying!

\begin{figure}[]
\centering
  \includegraphics[width=6cm]{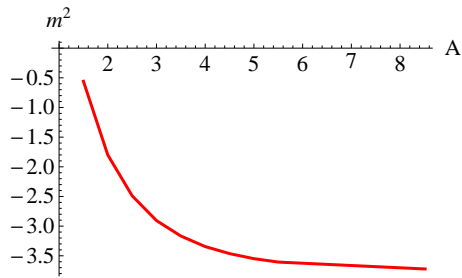}
            \caption{
           {The mass squared of the inflaton mode in the model with a step dilaton function (\ref{beta})
           against the step height $A$.}
           } \label{mass}
\end{figure}

In the Figure \ref{mass}, we show $m^2$ of this inflaton mode against the parameter $A$ in (\ref{beta}). We
change $\lambda$ in the dilaton each time to keep the vacuum energy constant. The plot again emphasizes that
the best inflationary models are when we tune $A$ to its critical value.

In principle, there are an infinite number of higher $m^2$ eigenvalue modes defined by 
(\ref{EOMR}), apart from this lowest $m^2$ we have found above (Strum Louiville theory enforces that the lightest state is that with no nodes which we have found). If these all have positive $m^2$, and hence
negative $\alpha$, then their
contribution to the evolution will die away exponentially fast with time (\ref{Tsol}). Could there be other sub-leading
negative $m^2$ states though? To attempt to answer this we numerically solve (\ref{EOMR}) starting with $R(\rho_c)=0$
and seek modes that fall to zero at large $\rho$. The resulting eigenvalues are $m^2 = -2.907, 11.18, 34.49$. The unstable mode we have already discussed is the only negative mass squared state. Of course
our IR boundary condition is somewhat ad hoc but the conclusion that there is just a single unstable mode seems robust
to its variation.

\begin{figure}[]
\centering
\includegraphics[width=6.5cm]{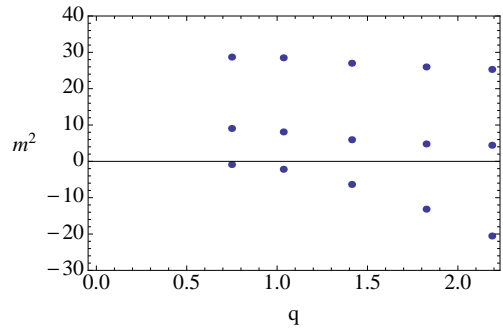}
\caption{{ The lightest three mesons' $m^2$ against the parameter $q$ in (\ref{beta2}) determined from (\ref{EOMR}) with $r_c=0.0127$.} }
\label{mvsA}
\end{figure}

So far our analysis of this subsection has been for our first ansatz for $\beta$ (\ref{beta}). The same phenomena apply
for the more sophisticated ansatz in (\ref{beta2}). To show there is just a  single inflaton state we solve 
(\ref{EOMR}) starting with $R(\rho_c)=0$
and seeking modes that fall to zero at large $\rho$, but now with $\beta$ from (\ref{beta2}). The results are $m^2 = -10, 2.495, 20.08$, when $B=10, q=1, r_c=0.0127$. 
We plot the lightest three mesons' $m^2$ versus the parameter $q$ of (\ref{beta2}) in Figure \ref{mvsA}. There is a single negative
mass squared state for values of $q$ above the transition point as advertised. Again this result is robust to changes in
the IR cut off.

\subsection{A single effective Inflaton}

Let us stress the conclusion of the analysis of the previous subsection. For the majority of the
roll time in this strongly coupled gauge theory the dynamics is dominated by a single negative
mass squared, linearized mode. Higher modes have positive mass and their contributions to the
evolution are exponentially suppressed with time.

With a given solution $R(\rho)$ the Lagrangian (\ref{Lag0}) boils down to
a standard scalar inflaton field:
\begin{eqnarray}
  && \call_{(0)} =   \sqrt{-g} \left(\half\dot\phi^2 - \half m^2 \phi^2\right) \,, \label{effec}
\end{eqnarray}
where $\sqrt{-g} = a(t)^3$ and
\begin{eqnarray}
  \phi(t) \equiv T(t) \sqrt{\int d \rho \frac{\b_0}{\rho} R(\rho)^2} \ . \label{phi}
\end{eqnarray}
Here we used  (\ref{m2}), and discarded the boundary term.

Since the model is simply a single inflaton model the phenomenology for the bi- and tri-spectra are 
already well explored. For a model that provides sufficient e-folds of inflation any signal will
be well below that one could hope to observe in envisaged experiments \cite{Chen:2010xka}.

\section{The $\eta$ problem}

Inflation models are frequently beset by the so called $\eta$ problem \cite{Copeland:1994vg}. One sets up a
scalar potential that is tuned sufficiently flat to generate slow roll inflation. The resulting
expansion term in the metric though can feedback into the inflaton potential generating
a mass of order the vacuum energy H and lead to a loss of slow roll. The only solution is to
further fine tune the potential so it remains flat in the presence of the back reaction.
A similar problem exists in DBI inflation too as pointed out in \cite{Chen:2008hz} and that mechanism
is similar to how the $\eta$ problem manifests in our set up.

So far we have treated the D7 brane as a probe and neglected the effect of its back-reaction
on the AdS metric, or equivalently the effect of the quarks on
the ${\cal N}=4$ gauge fields. At next order the D7's contribution to the vacuum energy
should deform the AdS$_5$ geometry. This back-reaction in principle will induce the breaking
of the SO(6) symmetry in the background geometry since the D7 does not fill the whole S$^5$.
That breaking has been studied in \cite{Polchinski,Kirsch:2005uy,Erdmenger:2011bw,Bigazzi:2011db} and induces a weak running of the gauge coupling
and a Landau pole. These changes are most important in the UV and reflect the true nature of
the ${\cal N}=2$ gauge theory underlying the construction. Our philosophy here is to use the IR
of this model, with a phenomenological dilaton profile, to represent a wider class of gauge
theories. For this reason we will neglect this particular aspect of the back-reaction.
Instead we will concentrate here on the  presence of the non-zero vacuum energy $\Delta V$.

The non-zero vacuum energy generates an inflating background geometry for the ${\cal N}=4$ gauge
fields to live in. That breaking of conformality should change the glue dynamics and hence change the background
AdS geometry. A geometry describing the ${\cal N}=4$ fields in an expanding background has been found in
\cite{DeWolfe:1999cp,Chen:2008hz} and is given by
\beq ds^2 = h(r)^2 (-dt^2 + a(t)^2 dx_3^2) + {dr^2 \over h(r)^2} + R^2 d\Omega_5^2 \,, \eeq
where
\beq h = \sqrt{ {r^2 \over R^2} - H^2 R^2} \,. \eeq
Note that at large $r$ the geometry simply returns to AdS$_5$. It is helpful in our
context to make the change of variables
\beq r = {H R^2 \over 2} \tilde{r} \left( 1 + {1 \over \tilde{r}^2} \right),\quad \tilde{r} = \frac{r+\sqrt{r^2 - H^2 R^4}}{H R^2} \,. \eeq
The geometry is now of the form
\beq ds^2 = h(\tilde{r})^2 (-dt^2 + a(t)^2 dx_3^2) + {R^2 \over \tilde{r}^2} (d \tilde{\rho}^2 + \tilde{\rho}^2 d\Omega_3^2 +
d \tilde{w}_5^2 + \tilde{w}_6^2 ) \,. \eeq
If we consider static embeddings of the probe D7 the DBI action is given by 
(\ref{DBI})
\beq \frac{{\cal L}}{ \tilde{T}_7 R^4 } = -\int d \tilde{\rho} \beta \frac{h^4}{\tilde{r}^4} \tilde{\rho}^3 \sqrt{1 + \tilde{L}^{'2} + \frac{\dot{L}^2}{h^2}  } \,. \eeq
Clearly the factor of $h^4$ changes the embeddings but we can see how to
change the running dilaton/coupling to return to the critical walking scenario solutions we used
above for inflation. We now require
\begin{align} 
\tilde{\beta} &\equiv \left(\frac{2}{H}\right)^4\beta (\tilde{r}) \frac{h^{4}}{\tilde{r}^4}  \\ 
&= \Big(A + 1 - A \tanh [(\tilde{r} - \lambda)] \Big)
 \left(1-\frac{1}{\tilde{r}^2}\right)^4 \,,
\end{align}
where we normalized $\beta$ so that $\tilde{\beta} \ra 1$ as $\tilde{r} \ra \infty$. 
If we start with the initial curved embedding lying outside $\tilde{r} \gg 1$ then $\tilde{\b} \sim \b$ and $h \sim \tilde{r}$.
\begin{figure}[]
\centering
  \includegraphics[width=6cm]{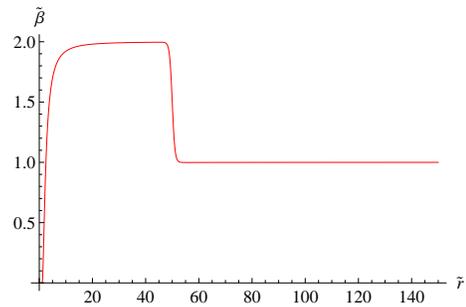}
            \caption{ $\tilde{\beta}$
           {  }
           } \label{modify}
\end{figure}

We will then have the same critical values of $A$ and $\lambda$ in (\ref{beta}) or $q$ and $B$ in (\ref{beta2}) . At those critical points the energy
difference between the symmetric flat embedding and the symmetry breaking curved embedding again falls
to zero leaving a flat potential between.

We show an example plot of $\tilde{\beta}$ in the $\tilde{r}$ coordinates in Figure \ref{modify} for the case of the step function $\beta$ (\ref{beta}) - here we pick 
some representative parameters with $\Delta V < \lambda$. 
As shown the running has only very small variation from the previous case
around the scale $\lambda$ but then large deviation at the scale $H << \lambda$. The required form around the scale
$H$ looks very peculiar and it is hard to imagine how it could be achieved in a gauge theory. In fact this is not
necessary though. The potential is rather flat for all smooth configurations between the flat embedding and the true
vacuum embedding. To achieve inflation we do not need to begin with a completely flat embedding that penetrates
down to $r=0$. Instead we can have some initial condition that lies in the range $H < L < \lambda$ so that the
brane only ever sees $\tilde{\beta}$ where it lies close to the ansatz for $\beta$ in
previous sections. Such small changes from our original ansatz seem no more or less arbitrary than the
original ansatz. The conclusion that walking dynamics could be responsible for inflation remains even including
the back-reaction of the $\eta$ problem.

Of course we must stress here that we have not in any sense solved the fine tuning problems associated with
generating inflation or solving the $\eta$ problem. These tunings are necessarily still present in our choice
of dilaton profile $\beta$ (and indeed we have not tried to backreact $\beta$ on the geometry either).
We hope though that we have recast those problems in an interesting fashion
that teaches us about the properties of a strongly coupled gauge theory that might cause inflation.

\acknowledgements

The authors thank Xingang Chen for discussions.
NE is grateful for the support of an STFC rolling grant. JF is grateful for University of Southampton Scholarships. KK acknowledges support via an NWO Vici grant of K. Skenderis. This work is part of the research program of the Stichting voor Fundamenteel Onderzoek der Materie (FOM), which is financially supported by the Nederlandse Organisatie voor Wetenschappelijk Onderzoek (NWO).

\end{document}